\documentclass[prd,aps,showpacs,nofootinbib,preprint,tightenlines,superscriptaddress]{revtex4}
\usepackage{epsfig,color}
\usepackage[latin1]{inputenc}
\usepackage{float,amsmath,slashed}
\usepackage{graphicx}

\newcommand{\beq}{\begin{equation}}
\newcommand{\eeq}{\end{equation}}
\newcommand{\bqa}{\begin{eqnarray}}
\newcommand{\eqa}{\end{eqnarray}}

\def\lsim{\mathrel{\rlap{\lower4pt\hbox{\hskip1pt$\sim$}}
    \raise1pt\hbox{$<$}}}                
\def\gsim{\mathrel{\rlap{\lower4pt\hbox{\hskip1pt$\sim$}}
    \raise1pt\hbox{$>$}}}                

\begin{document}

\title{Memory effects in radiative jet energy loss}

\author{Frank Michler}
\affiliation{Institut f\"ur Theoretische Physik \\
  Johann Wolfgang Goethe - Universit\"at Frankfurt \\
  Max-von-Laue-Stra\ss{}e~1,
  D-60438 Frankfurt am Main, Germany \vspace*{5mm}}

\author{Bj\"orn Schenke}
\affiliation{Department of Physics, McGill University, Montreal, Quebec, H3A\,2T8, Canada}

\author{Carsten Greiner}
\affiliation{Institut f\"ur Theoretische Physik \\
  Johann Wolfgang Goethe - Universit\"at Frankfurt \\
  Max-von-Laue-Stra\ss{}e~1,
  D-60438 Frankfurt am Main, Germany \vspace*{5mm}}

\begin{abstract}
In heavy-ion collisions the created quark-gluon plasma forms a quickly evolving background, leading to a time dependent radiative behavior of high momentum partons traversing the medium. We use the Schwinger Keldysh formalism to describe the jet evolution as a non-equilibrium process including the Landau-Pomeranschuk-Migdal effect. Concentrating on photon emission, a comparison of our results to a quasistatic calculation shows good agreement, leading to the conclusion that the radiative behavior follows the changes in the medium almost instantaneously.
\end{abstract}
\maketitle
\newpage

\small

\section{Introduction}
Relativistic heavy-ion collision experiments, as performed at the Schwerionen Synchrotron (SIS) at the Gesellschaft für Schwerionenforschung (GSI), the Super Proton Synchrotron (SPS) at CERN, the Relativistic Heavy Ion Collider (RHIC) at Brookhaven National Laboratory (BNL) and the future Large Hadron Collider (LHC) at CERN, allow for studying strongly interacting matter under extreme conditions. One main objective of these experiments is the creation and exploration of a new state of matter of deconfined quarks and gluons, the so called quark-gluon plasma (QGP). The theory of quantum chromodynamics (QCD) predicts that it is created at high densities and temperatures which occur during heavy-ion collisions at high energy.

The lifetime of the QGP is of the order of up to $5$ fm/c. After that, it transforms into a gas of hadrons. Thus experiments cannot directly access the quark-gluon plasma, which makes the determination of the properties of this state difficult. Therefore, it is important to find theoretical signatures which provide a distinction between a hadron gas and a quark-gluon plasma. Furthermore one has to find experimental observables from which one can draw conclusions on these theoretical signatures \cite{MN:2006,Yag:2005}. An important category of these observables are hard probes, i.e., partons with high transverse momenta produced during the early stage of a heavy-ion collision. These so called parton jets are significantly modified as they traverse the QGP. The properties of the medium can then be inferred by comparing the resulting hadron spectra to those for a simple p+p collision where medium effects are absent. Upon comparison to theoretical predictions for a hadron gas and a QGP one can draw conclusions on the latter one. The output of RHIC experiments \cite{Adams:2005dq,Reygers:2005ua,Adler:2003qi} shows that strong jet quenching occurs in heavy-ion collisions. The yields of high $p_{T}$ hadrons are suppressed with respect to the p+p reference by a factor of five.

Over the past decades, several theoretical formalisms \cite{Baier:1996kr,Gyulassy:2000er,Kovner:2003zj,Zakharov:2007pj,Wang:2001if,Arnold:2001ms,Arnold:2002ja}
have been established to describe energy loss due to bremsstrahlung processes. These radiative processes turned out to contribute strongly to the energy loss in a partonic medium \cite{Bjorken:1982tu,Zakharov:2007pj,Renk:2007id,Fochler:2008ts}, while collisional energy loss may not be negligible \cite{Mustafa:2003vh,Mustafa:2004dr,Adil:2006ei,Wicks:2007mk,Majumder:2008zg,Wang:2006qr,Qin:2007rn,Schenke:2009ik}.

Analogously to Bremsstrahlung in QED where photons are emitted, a parton which moves through the medium is expected to emit gluons. Here it is of particular interest how the standard pQFT cross section is modified by multiple scattering of the initial parton \cite{BD:1964,PS:1995}. In the framework of QED, this has already been studied by Landau, Pomeranschuk \cite{LP:1953} and Migdal \cite{Mig:1956}, who discovered a destructive interference phenomenon of the radiation from different scattering points. It occurs at small transverse photon momenta and thus leads to a reduction of the pQFT result in that range. The general arguments leading to this so called Landau-Pomeranschuk-Migdal (LPM) effect indicate a comparable behavior of radiative processes in QCD. Calculations by Gyulassy and Wang \cite{GW:1994} as well as subsequent works \cite{Baier:1994bd,Baier:1996kr,Arnold:2001ms,Arnold:2002ja} have indeed verified this.

Furthermore, the jet traverses a dynamic medium that expands and cools down. Thus the density of scattering centers is reduced which makes the radiative behavior of the jet time dependent as well. In this work, we study whether it adjusts to changes of the medium instantaneously (Markovian case) or if certain memory effects to be described below are of importance.

The non-equilibrium dynamics of heavy-ion collisions has always been a major motivation for investigations on non-equilibrium quantum field theory. In this context, the role of memory effects during time evolution has first been investigated by \cite{Greiner:1994xm,Kohler:1995zz}. In these works, non Markovian effects have been found to appear in the collision term of quantum transport equations. Later on, these effects have also been observed in the time evolution of so called disordered chiral condensates \cite{Xu:1999aq} as this evolution is governed by dissipative and non Markovian stochastic equations of motion. Furthermore, memory effects also occur in the thermalization of out of equilibrium $\phi^4$-theory \cite{Juchem:2004cs,Juchem:2003bi} as well as in the dilepton production from non equilibrated hot hadronic matter \cite{SG:2005,Schenke:2006uh}. In principle, the underlying field theoretical description always leads to equations of motion being non-local in time. This indicates that memory effects might also be relevant for the radiative energy loss of a jet transversing a dynamic QGP.

In this work, we use the non-equilibrium Green's function formalism first developed by Keldysh and Schwinger \cite{Schw61,Kel64,Kel65,Cra68} to study this non-equilibrium phenomenon. We approach the problem by first considering photon emission. In particular, our work has two main goals. First, we want to find the timescales in which the photon production rate adjusts to changes in the medium and determine their dependence on the momenta of the emitted photons. Second, we dynamically calculate the energy loss of the fermion jet over the time interval in which the plasma expands and cools down. Comparing our results to a quasistatic calculation determines the possible importance of memory effects for in-medium photon emission.

This paper is organized as follows. In Section II, we derive an expression which relates the photon rate with the current-current-correlator of the emitting system. This expression is non-local in time and hence accounts for the finite memory of the system. As source particle, we consider an ultrarelativistic fermion jet. In Section III, we present numerical results. After a short description of how the time dependencies have been implemented, we determine how the memory times of the partial photon rates depend on the momenta of the emitted photons. After that, we calculate the memory times for the total photon number and the total radiation power. Finally, we consider the energy loss of the fermion jet over the time interval in which the plasma expands and cools down. In Section IV, we close with a short summary of our results and an outlook to possible future research projects. We also touch on the differences between the QED and QCD frameworks.

\section{Non equilibrium photon production}
Since we concentrate on the investigation of time dependencies, it is sufficient to consider a system which is {\em spatially homogeneous}. In this case, the inclusive photon spectrum reads
\begin{equation}
 \label{eq:photonspectrum}
 \frac{d^{6}n_{\gamma}}{d^{3}xd^{3}k}(t) = \frac{1}{(2\pi)^{3}}\sum_{\lambda}
                                           \left\langle \hat{N}_{\lambda,H}(\vec{k},t)\right\rangle
                                         = \frac{1}{(2\pi)^{3}}\sum_{\lambda}
                                           \left\langle \hat{a}_{\lambda,H}^{\dagger}(\vec{k},t)
                                           \hat{a}_{\lambda,H}(\vec{k},t)\right\rangle \ ,
\end{equation}
where $\hat{N}_{\lambda,H}(\vec{k},t)$ denotes the number operator for photons with polarizations $\lambda$ and momentum $\vec{k}$.  Note that operators are indicated by $\hat{\cdot}$. They are expressed in the Heisenberg picture which is denoted by the subscript 'H'. As we consider a system which is out of equilibrium, the average is taken with respect to the initial density matrix. The creation and annihilation operators can be expressed in terms of the electromagnetic field operator and its time derivative:
\begin{subequations}
 \label{eq:operators}
 \begin{eqnarray}
  \hat{a}_{\lambda,H}(\vec{k},t)           & = & -\frac{\epsilon^{(\lambda)}(\vec{k})}{\sqrt{2k}}
                                                 \cdot\left[i\dot{\hat{A}}_{H}(\vec{k},t)+k\hat{A}_{H}(\vec{k},t)\right]
		           	                 \label{eq:operators1} \ , \\
  \hat{a}^{\dagger}_{\lambda,H}(\vec{k},t) & = & \frac{\epsilon^{(\lambda)\,*}(\vec{k})}{\sqrt{2k}}
                                                 \cdot\left[i\dot{\hat{A}}^{\dagger}_{H}(\vec{k},t)-k 
				 	         \hat{A}^{\dagger}_{H}(\vec{k},t)\right]
		         	                 \label{eq:operators2} \ .
 \end{eqnarray}
\end{subequations}
Upon insertion of (\ref{eq:operators1}) and (\ref{eq:operators2}) into (\ref{eq:photonspectrum}), we find
\begin{equation}
 \label{eq:photonspectrum_prop}
 2k\frac{d^{6}n_{\gamma}}{d^{3}xd^{3}k}(t) = \frac{1}{(2\pi)^3}\gamma^{\mu\nu}(\vec{k})\left.
					     \left(\partial_{t}\partial_{t'}+ik(\partial_{t}-\partial_{t'})+k^2\right)
					     iD^{<}_{\nu\mu}(\vec{k},t',t)\right|_{t=t'} \ , 
\end{equation}
where we have introduced the photon propagator $iD^{<}_{\nu\mu}(\vec{k},t',t)$ which is given by $$iD^{<}_{\nu\mu}(\vec{k},t',t)=\langle\hat{A}_{\mu,H}^{\dagger}(\vec{k},t)\hat{A}_{\nu,H}(\vec{k},t')\rangle.$$ $\gamma^{\mu\nu}(k)$ denotes the polarization tensor which reads:
\begin{equation}
\gamma^{\mu\nu}(\vec{k}) = \sum_{\lambda=1}^{2}\epsilon_{\mu}^{(\lambda)*}(\vec{k})
                               \epsilon_{\nu}^{(\lambda)}(\vec{k})
                         = -g_{\mu\nu}-\kappa_{\mu}\kappa_{\nu}+n_{\mu}n_{\nu} \ .
\end{equation}
Here we have introduced the timelike unit vector $n^{\mu}=(1,0,0,0)$ as well as the spacelike unit vector:
\begin{equation}
 \kappa^{\mu}=\frac{k^{\mu}-(k\cdot n)n^{\mu}}{\sqrt{(k\cdot n)^2-k^2}} \ .
\end{equation}
Up to here, our calculations have basically followed \cite{Serreau:2003wr}. In the next step, we have to calculate the photon propagator. Instead of employing a strictly perturbative analysis as carried out in \cite{Serreau:2003wr}, we evaluate $D^{<}_{\nu\mu}(\vec{k},t',t)$ using a generalized fluctuation-dissipation-relation \cite{Danielewicz:1982kk,Greiner:1998vd,SG:2005}
\begin{equation}
 \label{eq:fdr_phot}
 D^{<}_{\nu\mu}(\vec{k},t',t) = \int_{-\infty}^{t'}\int_{-\infty}^{t}dt_1\,dt_2
                                D^{R}_{\nu\alpha}(\vec{k},t',t_1)\Pi^{<}_{\alpha\beta}(\vec{k},t_1,t_2)
				D^{A}_{\beta\mu}(\vec{k},t_2,t) \ ,
\end{equation}
with the retarded and advanced photon propagators $D^{R/A}$ and the photon self energy $i\Pi^{<}$. Since photons are expected to leave the medium undisturbed, $D^{R/A}$ can be approximated by the free propagators $D^{R/A}_{0}$:
\begin{subequations}
 \label{eq:free} 
 \begin{eqnarray}
  D^{R}_{\mu\nu,0}(\vec{k},t,t') & = & -\frac{1}{k}\sin\left(k(t-t')\right)\Theta(t-t')g_{\mu\nu}
                                      \label{eq:free1} \ , \\
  D^{A}_{\mu\nu,0}(\vec{k},t,t') & = & \frac{1}{k}\sin\left(k(t'-t)\right)\Theta(t'-t)g_{\mu\nu} 
                                      \label{eq:free2} \, .
 \end{eqnarray}
\end{subequations}
Upon insertion of (\ref{eq:free1})-(\ref{eq:free1}) into (\ref{eq:photonspectrum_prop}) and taking the time derivative, we obtain for the photon production rate at time $t$:
\begin{equation}
 \label{eq:photonrate}
 2k\frac{d^{7}n_{\gamma}}{d^{4}xd^{3}k}(t) = \frac{1}{(2\pi)^3}\gamma^{\mu\nu}(\vec{k})\int_{-\infty}^{t}
                                             du\cdot 2\text{Re}\left[ i\Pi^{<}_{\mu\nu}(\vec{k},t,u)
				  	     e^{ik(t-u)}\right] \ .
\end{equation}
This expression directly relates the photon production rate at a given time $t$ to the full history of the emitting system described by the current-current-correlator. This non-locality in time accounts for possible memory effects. Equation (\ref{eq:photonrate}) is a generalization of the corresponding equilibrium formula \cite{McLerran:1984ay,Gale:1987ki}, which it reproduces for stationary systems.

In the case of photon emission from a fermion jet moving through hot and dense matter, $i\Pi^{<}_{\mu\nu}(\vec{k},t,u)$, the self energy emerging from the fermion-photon interaction, is given by the irreducible part of the current-current-correlator in coordinate space
\begin{eqnarray}
 \label{eq:pse}
 i\Pi_{\mu\nu}(x,x') & = & \left\langle T_{C}\left(\hat{j}_{\mu,H}(x)\hat{j}^{\dagger}_{\nu,H}(x')
                           \right)\right\rangle_{\text{irred}} \nonumber \\
		     & = & \left\langle T_{C}\left[\exp\left(-i\slashed{\int}_{t_{0}}^{t_{0}} dt_{1}
		           \hat{H}_{I}(t_{1})\right)\hat{j}_{\mu,I}(x)\hat{j}^{\dagger}_{\nu,I}(x')\right]
			   \right\rangle_{\text{irred}}
\end{eqnarray}
where $\slashed{\int}_{t_{0}}^{t_{0}}$ describes the Keldysh countour integral starting and ending at time $t_0$ and running above the largest time argument of $i\Pi_{\mu\nu}(x,x')$. $T_C$ implies time ordering of the operators along the contour and $\hat{H}_{I}(t_{1})$ denotes the interaction part of the QED-Hamiltonian
\begin{equation}
 \hat{H}_{I}(t_{1}) = \int d^{3}x_{1}\hat{j}_{\mu,I}(x_{1})\hat{A}^{\mu}_{I}(x_{1}) \ .
\end{equation}
The subscript 'I' denotes that the operators are expressed in the interaction picture. In order to keep the notation more shorthand, we shall omit this subscript from now on. The current operators $\hat{j}_{\mu}(x)$ and $\hat{j}^{\dagger}_{\nu}(x')$ are given by
\begin{subequations}
 \label{eq:currentoperators}
 \begin{eqnarray}
  \hat{j}_{\mu}(x)            & = & e\hat{\bar{\psi}}(x)\gamma_{\mu}\hat{\psi}(x) 
                                    \label{eq:currentoperators1} \ , \\
  \hat{j}^{\dagger}_{\mu}(x') & = & e\hat{\bar{\psi}}(x')\gamma_{\mu}\hat{\psi}(x') 
                                    \label{eq:currentoperators2} \, .
 \end{eqnarray}
\end{subequations}
Irreducibility means that one includes only those diagrams which cannot be separated by cutting a single fermion line. Since the expectation value of an odd product of operators always vanishes, we can introduce an effective interaction operator
\begin{equation}
 \hat{V}_{\text{eff}} = \int d^{4}{x}_{1}\int\,d^{4}x_{2}\,\hat{j}_{\alpha}(x_{1})\hat{A}^{\alpha}(x_{1})
	                \hat{j}_{\beta}(x_{2})\hat{A}^{\beta}(x_{2}) \ ,
\end{equation}
and rewrite Equation (\ref{eq:pse}) as follows:
\begin{equation}
 \label{eq:pse_re}
 i\Pi_{\mu\nu}(x,x') = \sum_{n=0}^{\infty}\frac{(-1)^{n}}{(2n)!}\left\langle T_{C}\left[\hat{V}_{\text{eff}}^{n}  
		       \hat{j}_{\mu}(x)\hat{j}^{\dagger}_{\nu}(x')\right]\right\rangle_{\text{irred}}
\end{equation}
Next, we apply a finite temperature Wick decomposition \cite{Danielewicz:1982kk} to each term of (\ref{eq:pse_re}). In order to avoid double counting, we have to take into account only those diagrams which represent real interactions and are not already included in the full fermion propagators. Since the $n$-th order term of (\ref{eq:pse_re}) formally scales with $\alpha_{e}^{n+1}$, it is sufficient for our purpose to consider the $n=0$ term only, which leads to the standard one-loop-approximation of Eq. (\ref{eq:pse})
\begin{eqnarray}
 i\Pi_{\mu\nu}(x,x') & = & \left\langle T_{C}\left[\hat{j}_{\mu}(x)\hat{j}^{\dagger}_{\nu}(x')
                           \right]\right\rangle \nonumber \\  
                     & = & \Theta_{C}(t,t')\left\langle \hat{j}_{\mu}(x)\hat{j}^{\dagger}_{\nu}(x')
                           \right\rangle + \Theta_{C}(t',t)\left\langle \hat{j}^{\dagger}_{\nu}(x')
                           \hat{j}_{\mu}(x)\right\rangle \nonumber \\
	             & = & \Theta_{C}(t,t')i\Pi^{>}_{\mu\nu}(x,x') + 
		           \Theta_{C}(t',t)i\Pi^{<}_{\mu\nu}(x,x') \ ,
\end{eqnarray}
Here $\Theta_{C}$ denotes the step function generalized to the time contour $C$. $i\Pi^{>}_{\mu\nu}(x,x')$ and $i\Pi^{<}_{\mu\nu}(x,x')$ are given by
\begin{subequations}
 \label{eq:polindentities}
 \begin{eqnarray}
  i\Pi^{>}_{\mu\nu}(x,x') & = & \left\langle \hat{j}_{\mu}(x)\hat{j}^{\dagger}_{\nu}(x')
                                \right\rangle \ , \label{eq:polindentities1} \\
  i\Pi^{<}_{\mu\nu}(x,x') & = & \left\langle \hat{j}^{\dagger}_{\nu}(x')\hat{j}_{\mu}(x)
                                \right\rangle \ . \label{eq:polindentities2}
 \end{eqnarray}
\end{subequations}
Insertion of (\ref{eq:currentoperators1})-(\ref{eq:currentoperators2}) into (\ref{eq:polindentities2}) and application of the finite temperature Wick theorem yields:
\begin{equation}
 \label{eq:poleval}
 i\Pi^{<}_{\mu\nu}(x,x') = e^2\cdot \text{Tr}\left\lbrace\gamma_{\mu}S^{<}_{F}(x,x') 
			   \gamma_{\nu}S^{>}_{F}(x',x) \right\rbrace \ . 
\end{equation}
Here $S^{<}(x,x')$ and $S^{>}(x',x)$ are the fermion propagators which are given by
\begin{subequations}
 \begin{eqnarray}
  -iS^{>}_{\alpha\beta}(x,x') & = & \left\langle \hat{\bar{\psi}}_{\beta}(x')\hat{\psi}_{\alpha}(x)
                                    \right\rangle \ , \\
   iS^{<}_{\alpha\beta}(x',x) & = & \left\langle \hat{\psi}_{\alpha}(x')\hat{\bar{\psi}}_{\beta}(x)
                                    \right\rangle \ .
 \end{eqnarray}
\end{subequations}
For a spatially homogeneous system, $i\Pi^{<}_{\mu\nu}(\vec{k},t,u)$ reads
\begin{equation}
 \label{eq:polmomentum}
  i\Pi^{<}_{\mu\nu}(\vec{k},t,u) = e^{2}\int\frac{d^{3}p}{(2\pi)^3}Tr\left\lbrace
                                   \gamma_{\mu}S^{<}_{F}(\vec{p}+\vec{k},t,u) 
 			           \gamma_{\nu}S^{>}_{F}(\vec{p},u,t)\right\rbrace.
\end{equation}
Assuming that the emitting system is in local equilibrium, the Kubo Martin Schwinger (KMS) relation \cite{kubo57,Schw61} yields for the propagators in frequency space
\begin{subequations}
 \label{eq:ft_prop}
  \begin{eqnarray}
   S^{<}_{F}(\vec{p}+\vec{k},E) & = & 2\pi i \,n_{F}(E)A_{F}(\vec{p}+\vec{k},E) 
                                      \label{eq:ft_prop1} \ , \\
   S^{>}_{F}(\vec{p},E)         & = & -2\pi i \,(1-n_{F}(E))A_{F}(\vec{p},E)
                                      \label{eq:ft_prop2} \ .
  \end{eqnarray}
\end{subequations}
Here $A_{F}(\vec{p},E)$ is the spectral function and $n_{F}(E)$ denotes the equilibrium Fermi distribution. Since we want to consider the radiation arising from a highly energetic fermion jet, we make the following ansatz. We associate $S^{<}(\vec{p}+\vec{k},E)$ with the ingoing particle and replace $n_{F}(E)$ by an effective distribution $f(E,\vec{p})$ which includes the fermion jet as one additional particle with fixed momentum $\vec{p'}$:
\begin{equation}
 \label{eq:jet}
 n_{F}(E) \rightarrow f(E,\vec{p}+\vec{k}) = n_{F}(E) + \frac{(2\pi)^{3}}{V}\delta^{(3)}(\vec{p}+\vec{k}-\vec{p'}) \ .
\end{equation}
Here $V$ denotes the volumes of the emitting system. Since we are only interested in the radiation from the fermion jet, we neglect the Fermi distribution 
$n_{F}(E)$, and $S^{<}(\vec{p},E)$ thus reduces to:
\begin{equation}
 \label{eq:smaller_re}
  S^{<}(\vec{p}+\vec{k},E) = 2\pi i\frac{(2\pi)^{3}}{V}\delta^{(3)}(\vec{p}+\vec{k}-\vec{p}^{'})A_{F}(\vec{p},E) \ .
\end{equation}
If we further associate $S^{>}(\vec{p},E)$ with the outgoing particle and assume that the mean occupation number of this state is negligibly small before the emission process, we can replace $1-n_{F}(E)$ by unity and obtain
\begin{equation}
 \label{eq:bigger_re}
  S^{>}(\vec{p},E) = -2\pi i\,A_{F}(\vec{p},E) \ .
\end{equation}
The latter assumption is legitimate as we consider a highly energetic fermion jet scattering into unoccupied states. For $A_{F}(\vec{p},E)$ we use the fermionic Breit Wigner structure as done in \cite{Henning:1995ww}. It generalizes the standard onshell energy momentum relation
\begin{equation}
 \label{eq:delta}
 A^{\text{free}}_{F}(\vec{p},E) = (E\gamma_{0}-\vec{\gamma}\cdot\vec{p}+m)\,\mathrm{sign}(E)\,\delta(E^2-\varepsilon_{\vec{p}}^{2}) \ ,
\end{equation}
to a broad distribution of interacting particles, which is described by a double Lorentzian curve with half maximum width $2\Gamma$:
\begin{equation}
 \label{eq:breitwigner}
 A_{F}(\vec{p},E) = \frac{\Gamma}{\pi}\frac{\gamma_{0}(E^{2}+\varepsilon_{\vec{p}}^{2}+\Gamma^{2})
                    -2E\vec{\gamma}\cdot\vec{p}+2Em}{(E^{2}-\varepsilon_{\vec{p}}^{2}-\Gamma^{2})^2+
		    4E^{2}\Gamma^{2}} \ .
\end{equation}
Here $\varepsilon_{\vec{p}}$ denotes the onshell energy for relativistic fermions. Due to the non zero spectral width parameter $\Gamma$, the Fourier transform of (\ref{eq:breitwigner}) into the time representation is exponentially damped. So the spectral width represents a finite lifetime of every excitation in the medium and can thus be associated with the microscopic scattering rate of the source particles. For $\Gamma\rightarrow 0$, one recovers the free spectral function $A^{\text{free}}_{F}(\vec{p},E)$ from (\ref{eq:breitwigner}).

For quarks in a QGP, the scattering width $\Gamma$ has contributions both from electromagnetic and strong interaction processes \cite{Henning:1995ww}:
\begin{equation}
 \Gamma(T) = \Gamma_{\text{em}}(T)+\Gamma_{\text{s}}(T) \ .
\end{equation}
For the temperatures $T$, which occur in a quark-gluon plasma, the contribution from strong interaction processes is proportional to the temperature times some constant of order 1, i.e., it basically behaves as $\Gamma_{\text{s}}(T)\approx T$ \cite{Henning:1995ww}. So $\Gamma$ may typically range from 0.1 to 1.0 GeV \cite{Peshier:2006mp,Xu:2007aa}. The electromagnetic contribution behaves as $\Gamma_{\text{em}}(T)\sim\alpha_{e}T$ and can be neglected.

Since we are only interested in the photon emission from a particle (not antiparticle), we take into account only the fermion component of (\ref{eq:breitwigner}), which has its peak at positive energies:
\begin{equation}
 A_{F}(\vec{p},E) = \frac{\Gamma}{2\pi\varepsilon_{\vec{p}}}\left\lbrace\frac{\gamma_{0}\varepsilon_{\vec{p}}
	            -\vec{\gamma}\cdot\vec{p}+m}{(E-\varepsilon_{\vec{p}})^{2}+\Gamma^2}\right\rbrace \ .
\end{equation}
Upon transformation of (\ref{eq:smaller_re}) and (\ref{eq:bigger_re}) into the time representation and evaluation of the trace in (\ref{eq:polmomentum}), we obtain for $i\Pi^{<}_{\mu\nu}(\vec{k},t,u)$:
\begin{eqnarray}
 \label{eq:pse_p}
 i\Pi^{<}_{\mu\nu}(\vec{k},t,u) & = & \frac{e^{2}V^{-1}}{\varepsilon_{\vec{p}}\,\varepsilon_{\vec{p}-\vec{k}}}
                                      \left\lbrace(p_{\mu}(p-k)_{\nu}+p_{\nu}(p-k)_{\mu}-g_{\mu\nu} 
				      \left[p\cdot(p-k)-m^2\right]\right\rbrace\nonumber \\
				&   & \times e^{-i(\varepsilon_{\vec{p}}-\varepsilon_{\vec{p}-\vec{k}})(t-u)}e^{-2\Gamma|t-u|}
\end{eqnarray}
Here we have introduced $p_{0}=\varepsilon_{\vec{p}}$ and $(p-k)_{0}=\varepsilon_{\vec{p}-\vec{k}}$. The superscript ' has been omitted for convenience. If we now insert (\ref{eq:pse_p}) into (\ref{eq:photonrate}), the volume factor $V$ drops out when integrating the photon spectrum over $d^{3}x$ and we obtain
\begin{eqnarray}
\label{eq:rate_final}
 k\frac{d^{4}n_{\gamma}}{dtd^{3}k}(t) & = & \frac{e^{2}}{(2\pi)^{3}}\gamma^{\mu\nu}(k)\,\frac{p_{\mu}(p-k)_{\nu}
                                            + p_{\nu}(p-k)_{\mu}-g_{\mu\nu}\left[p\cdot(p-k)-m^2\right]}
				            {\varepsilon_{\vec{p}}\,\varepsilon_{\vec{p}-\vec{k}}} \nonumber \\
			              &   & \times \int_{-\infty}^{t}du \mbox{ }e^{-2\Gamma|t-u|}
			                    \cos\left(\omega_{k}(t-u)\right) \ .    
\end{eqnarray}
where $\omega_{k}$ is given by
\begin{equation}
 \omega_{k}=\varepsilon_{\vec{p}}-\varepsilon_{\vec{p}-\vec{k}}-k \ . 
\end{equation}
It is helpful to rewrite (\ref{eq:rate_final}) in terms of $k_{\parallel}$ and $k_{\perp}$, which denote the components of the photon momentum parallel and perpendicular to the jet momentum respectively. We then carry out the integration over the azimuthal angle $\varphi_{k}$ and obtain
\begin{eqnarray}
\label{eq:rate_final_re}
 \frac{k}{k_{\perp}}\frac{d^{3}n_{\gamma}}{dtdk_{\parallel}dk_{\perp}}(t) 
      & = & \frac{\alpha_{e}}{\pi}\left[\left(1-\frac{k_{\parallel}^{2}}{k^{2}}\right)\left(1+
            \frac{p(p-k_{\parallel})-m^{2}}{\varepsilon_{\vec{p}}\varepsilon_{\vec{p}-\vec{k}}}\right)
	    +\left(2-\frac{k_{\perp}^{2}}{k^{2}}\right)\left(1-\frac{p(p-k_{\parallel})+m^{2}}
	    {\varepsilon_{\vec{p}}\varepsilon_{\vec{p}-\vec{k}}}\right)\right.\nonumber\\
      &   & \left.+\frac{2 k_{\perp}^2}{k^{2}}\frac{p\cdot k_{\parallel}}{\varepsilon_{\vec{p}}\varepsilon_{\vec{p}-\vec{k}}}\right]
            \times \int_{-\infty}^{t}du \mbox{ }e^{-2\Gamma|t-u|}\cos\left(\omega_{k}(t-u)\right) \ .
\end{eqnarray}
Expression (\ref{eq:rate_final_re}) allows for the investigation of the radiative behavior of a fermion jet in a non-equilibrated hot plasma. The expansion and cooling down of the plasma results in a time dependent scattering rate $\Gamma=\Gamma(T(\tau))\equiv\Gamma(\tau)$. Due to the time integral in (\ref{eq:rate_final_re}), the photon production rate at time $t$ does not only depend on $\Gamma$ at time $t$ but also on $\Gamma$ at earlier times. Therefore potential memory effects are covered by this expression.

For completeness, we briefly discuss the applicability of the one loop approximation. It corresponds to the direct radiation from the different scattering points whereas interference contributions are described by so called ladder diagrams \cite{Knoll:1995nz}. So interference contributions have not been taken into account explicitly. Nevertheless, the ladder diagrams can be summed up to an effective one loop diagram when assuming that a nonrelativistic velocity of the source particle is reduced by a constant fraction at each collision \cite{Knoll:1995nz}. In this case, the summation procedure effectively leads to a reduced scattering rate $\Gamma$. Therefore, we may consider our one loop diagram as an effective one where interference contributions and thus the LPM effect are already taken into account. Furthermore, the one loop approximation is formally the leading order contribution in the QED framework.

One easily verifies that (\ref{eq:rate_final_re}) cures the infrared divergence of the standard pQFT-result \cite{BD:1964,PS:1995}. This is because the problem of multiple scattering has been treated using full Green's functions with damping. The damping results from repeated scattering and leads to a suppression of the photon production rate for $\omega_{k}\ll 2\Gamma$. This range will from now on be referred to as LPM region. In the Bethe Heitler region, i.e., for $\omega_{k}\gg\Gamma$, our result reproduces (compare \cite{Knoll:1995nz} for the non relativistic case) the standard pQFT expression \cite{BD:1964,PS:1995}:
\begin{equation}
  \frac{d^{5}n}{d^{4}xdk}  = \frac{1}{2\pi^{2}}\cdot\frac{2\alpha_{e}\rho_{0}\Gamma}{k}\left\langle\
                             \frac{\gamma_{ij}(\vec{k})\beta_{i}\beta_{j}}{(1-\beta\cos{\theta})^{2}}\right\rangle_{\Omega_{k}} \ .
\end{equation}
Here $\left\langle\cdots\right\rangle_{\Omega_{k}}$ denotes the integration over the solid angle in $\vec{k}$-space. We shall give an explanation of how the suppression mechanism for $\omega_{k}\ll\Gamma$ comes about. The current of the source particle decorrelates within the time interval $\tau_{S}=1/(2\Gamma$). As this is the timescale over which the source particle 'remembers' a scattering, radiation induced by this scattering is only possible within this time interval. In the LPM region, however, the formation time $$\tau_{F}\simeq2\pi /\omega_{k}$$ of the photon is much larger than this decorrelation time. So the probability for the photon to be emitted within the decorrelation time interval is very small and the photon production rate is suppressed. This shows that the LPM effect is not purely an interference phenomenon. In fact, a perturbative treatment of interference contributions fails to cure the infrared divergence \cite{Fortmann:2005js}.

\section{Numerical Investigations and Results}
In order to investigate the radiative behavior of a fermion jet in a non equilibrated hot plasma, we simulate the evolving medium by varying the scattering rate $\Gamma$. The size of possible memory effects is most effectively extracted when changing $\Gamma$ linearly over different time intervals $\Delta t$. Since the integrand of (\ref{eq:rate_final_re}) depends on two different time arguments we have to decide on which time argument $\tau$ the scattering rate $\Gamma$ shall depend. It is possible to introduce $\Gamma=\Gamma(T(u))\equiv\Gamma(u)$ or $\Gamma=\Gamma\left(T\left(\frac{t+u}{2}\right)\right)\equiv\Gamma\left(\frac{t+u}{2}\right))$, i.e., $\Gamma$ can depend on the backward time $u$ or the center of mass time $\frac{t+u}{2}$. The choice $\Gamma=\Gamma(T(t))\equiv\Gamma(t)$ is also possible but it results in an entirely Markovian time evolution of (\ref{eq:rate_final_re}). For our numerical investigations, we consider a $p=20$ GeV fermion jet.
\begin{figure}[tbh]
 \begin{center}
  \includegraphics[height=5.0cm]{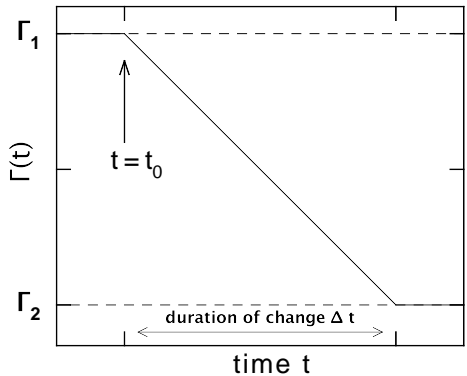}
  \caption{Linear change of $\Gamma$ over $\Delta t$.}
  \label{fig:linearchange}
 \end{center}
\end{figure}

As initial and final scattering rate, it is convenient to choose $\Gamma_{1}=1.0$ GeV and $\Gamma_{2}=0.3$ GeV, respectively. To find the timescales in which the photon rates adjust to changes in the medium and how these so called memory times depend on the momenta of the emitted photons, we consider the relative difference between the final (at $t\rightarrow \infty$) and an intermediate photon rate which we define as follows:
\begin{equation}
 r(\Delta t) = \left|\frac{d^{3}n(t\rightarrow\infty)-d^{3}n(t_{0}+\Delta t)}
               {d^{3}n(t\rightarrow\infty)-d^{3}n(t_{0})}\right|
\end{equation}
Here $t_{0}$ is the initial time at which $\Gamma$ starts to change and $t_{0}+\Delta t$ is the point of time where the change of $\Gamma$ is complete. In the Markovian case, the photon production rate at $t_{0}+\Delta t$ has already evolved to its final value so that $r(\Delta t)=0$. In the case of full memory, the photon production rate at $t_{0}+\Delta t$ still remains at its initial value and we have $r(\Delta t)=1$. In general, it has a finite value different from 1 and thus quantifies the memory effects for a given change duration $\Delta t$. Therefore, we are able to determine the memory times of the system by parametrizing $r(\Delta t)$ and adjusting the parameters to the result.

\begin{figure}[htb]
  \hfill
  \begin{minipage}[t]{.45\textwidth}
      \begin{center}
        \includegraphics[height=5cm]{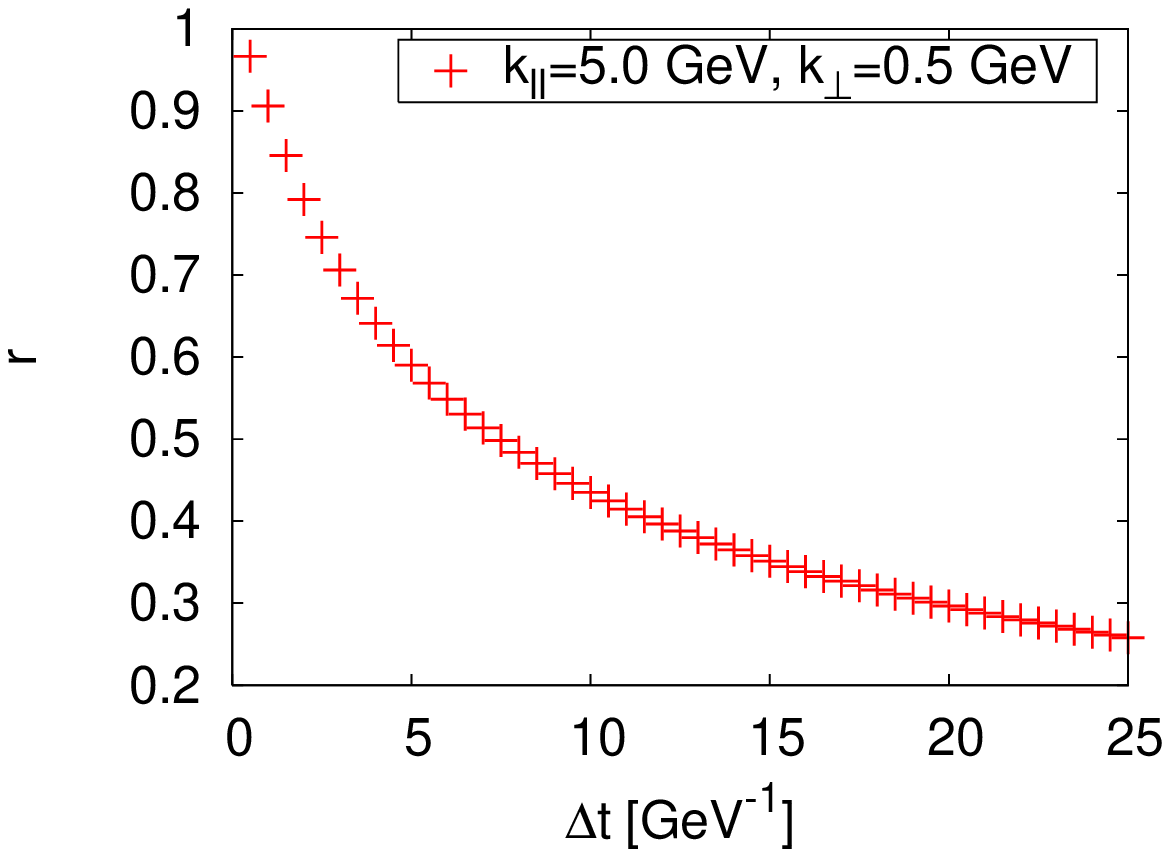}
        \caption{$r(\Delta t)$ for $k_{\parallel}=5.0$ GeV and $k_{\perp}=0.5$ GeV (LPM-region)}
        \label{fig:rd_lpm}
      \end{center}
  \end{minipage}
  \hfill
  \begin{minipage}[t]{.45\textwidth}
      \begin{center}
        \includegraphics[height=5cm]{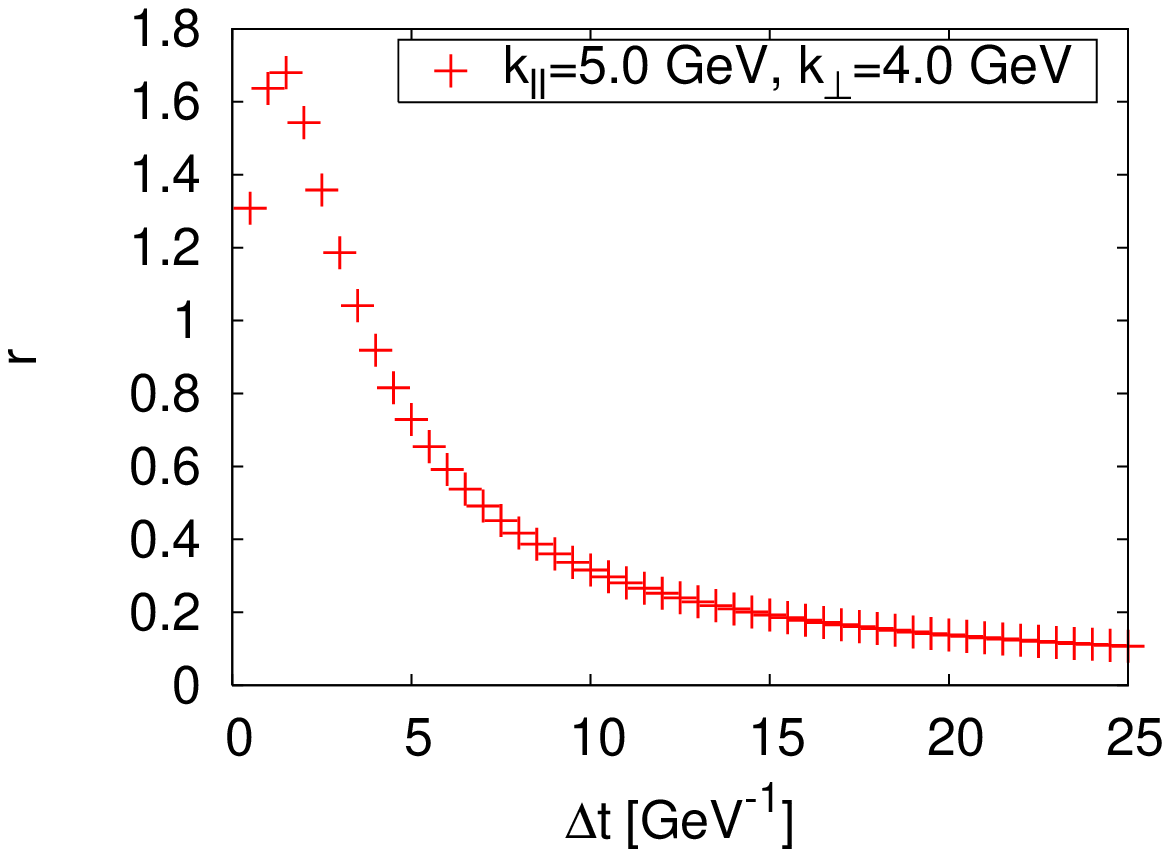}
        \caption{$r(\Delta t)$ for $k_{\parallel}=5.0$ GeV and $k_{\perp}=4.0$ GeV (BH-region)}
        \label{fig:rd_bh}
      \end{center}
  \end{minipage}
  \hfill
\end{figure}

Figs. \ref{fig:rd_lpm} and \ref{fig:rd_bh} show the relative difference $r(\Delta t)$. For $k_{\perp}\ll k_{\parallel}$ we find an exponential decay $r(\Delta t)\sim e^{-\Delta t/\tau}$ (Fig. \ref{fig:rd_lpm}) while for larger $k_{\perp}$ we have $r(\Delta t)\sim(\Delta t/\tau) e^{-\Delta t/\tau}$ (Fig. \ref{fig:rd_bh}). In both figures, $r(\Delta t)$ is shown for $\Gamma=\Gamma(u)$.

\begin{figure}[htb]
  \hfill
  \begin{minipage}[t]{.45\textwidth}
      \begin{center}
        \includegraphics[height=5cm]{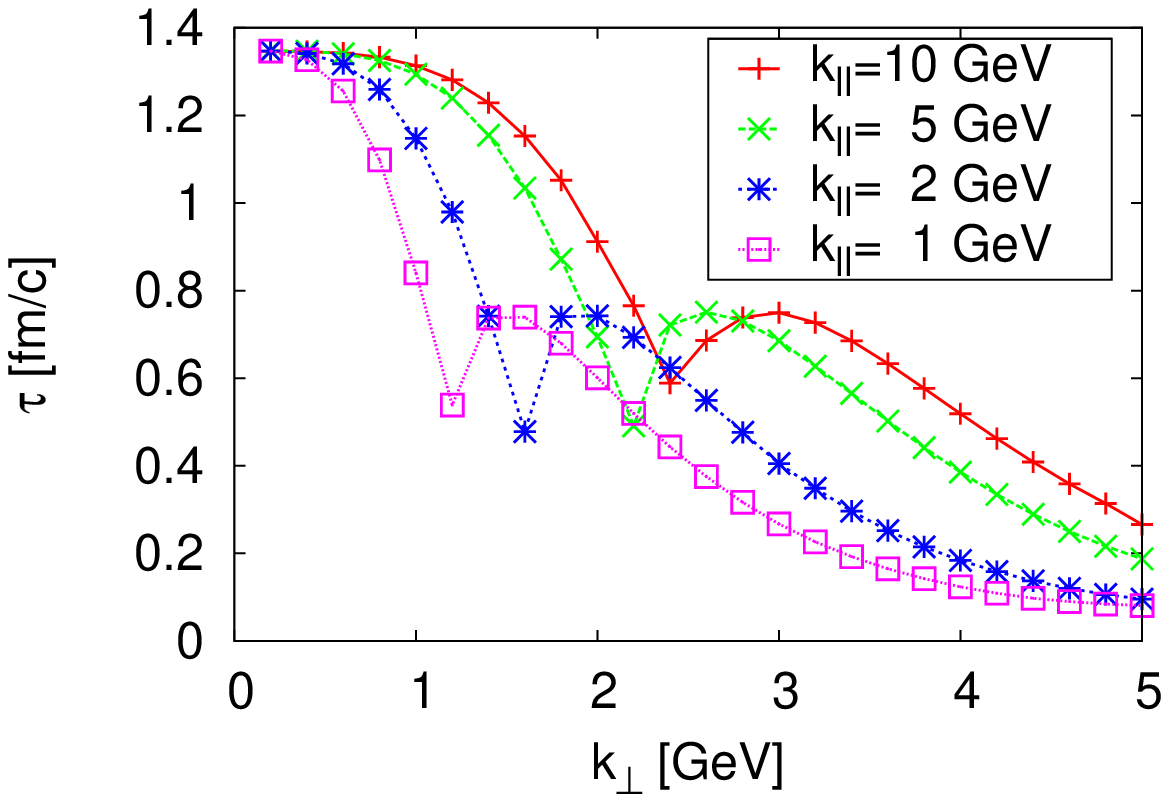}
        \caption{Memory times for $\Gamma=\Gamma(u)$}
        \label{fig:memory_1}
      \end{center}
  \end{minipage}
  \hfill
  \begin{minipage}[t]{.45\textwidth}
      \begin{center}
        \includegraphics[height=5cm]{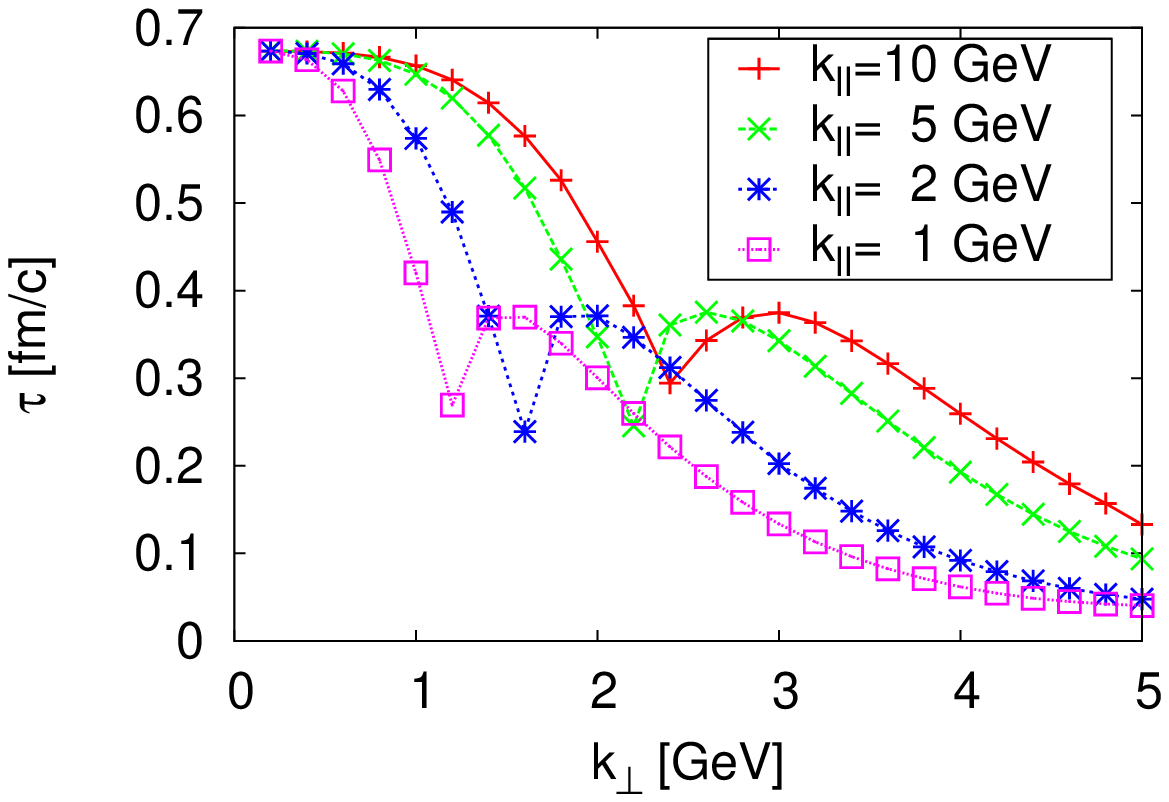}
        \caption{Memory times for $\Gamma=\Gamma[(t+u)/2]$}
        \label{fig:memory_2}
      \end{center}
  \end{minipage}
  \hfill
\end{figure}

So the following parametrization incorporates both cases
\begin{equation}
 \label{eq:fitfunction}
 r(\Delta t) = \left[A+B\frac{\Delta t}{\tau}\right]e^{-\Delta t/\tau} \ .
\end{equation}
The memory time is given by the parameter $\tau$, which describes how fast the system loses information about the initial state. Its values are shown in Figs. \ref{fig:memory_1} and \ref{fig:memory_2} for fixed $k_{\parallel}$ and varying $k_{\perp}$.
We find that the memory times become smaller at larger $k_\perp$, a behavior that can be understood when regarding the characteristic timescale $\langle\tau\rangle$ that contributes to the memory integral in (\ref{eq:rate_final_re})
\begin{equation}
 \label{eq:timescale}
 \langle\tau\rangle \simeq \text{Re}\left\lbrace\frac{\int_{-\infty}^{t}du(t-u)e^{-i\omega_{k}(t-u)}e^{-2\Gamma|t-u|}}
                           {\int_{-\infty}^{t}due^{-i\omega_{k}(t-u)}e^{-2\Gamma|t-u|}}\right\rbrace
		    =      \frac{2\Gamma}{\omega_{k}^{2}+4\Gamma^{2}}
\end{equation}
where $\omega_{k}$ reads in terms of $k_{\parallel}$ and $k_{\perp}$:
\begin{equation}
\omega_{k} = \omega_{k}(k_{\parallel},k_{\perp})
           = p-\sqrt{p^{2}+k^{2}_{\parallel}+k^{2}_{\perp}-2pk_{\parallel}}-\sqrt{k^{2}_{\parallel}+k^{2}_{\perp}}
\end{equation}
So the larger $k_{\perp}$, the shorter the contributing timescale to (\ref{eq:rate_final_re}) and hence the smaller the memory effects. In fact, if we compare Figs. \ref{fig:memory_1}-\ref{fig:memory_2} with (\ref{eq:timescale}) we see that the memory times essentially behave as the characteristic timescale $\langle\tau\rangle$ which can be understood as follows. For $k_{\perp}\ll k_{\parallel}$, i.e., for $\omega_{k}\ll 2\Gamma$, the decorrelation time $\tau_{S}=1/2\Gamma$ sets an upper bound for the memory time. That is why $\tau$ only depends on $\Gamma_{1}$ and $\Gamma_{2}$ but not on $k_{\parallel}$ in that range. One can indeed show \cite{Michler:2007} that the memory time behaves as $\tau\sim 1/\Gamma_{2}$ for $k_{\perp}\rightarrow 0$.

In the Bethe-Heitler regime ($k_{\perp}\gg k_{\parallel}$), the memory time behaves as $\tau\sim\tau_{F}/N$ where $N$ is the number of oscillations in the integrand of (\ref{eq:rate_final_re}) during the time interval $\tau_{S}$. So the expression that the memory time scales with is much smaller than the formation time $\tau_{F}$. Due to the oscillations only the fraction $1/N$ of the information from the timescale $\tau_{F}$ remains.

So the memory time $\tau$ is {\em not identical} to the formation time $\tau_{F}$ which one might expect. In particular, both quantities do not coincide with each other for $k_{\perp}\ll k_{\parallel}\ll p$. In this limit, $\tau_{F}$ can be approximated as:
\begin{equation}
 \label{eq:dokhshitzer}
  \tau_{F}\simeq\frac{k_{\parallel}}{k_{\perp}^2}=\frac{\gamma_{\text{eff}}(k_{\parallel},k_{\perp})}{k_{\perp}}
\end{equation}
which is also known as Dokshitzer's approximation \cite{Dok:1991}. One could already infer this from the decorrelation time $\tau_{S}$ which sets an upper bound for the memory time whereas (\ref{eq:dokhshitzer}) diverges for $k_{\perp}\rightarrow 0$. For completeness, we shall 
briefly point out how (\ref{eq:dokhshitzer}) comes about. In a reference frame where $k_{\parallel}=0$, the formation of a photon is given by $\tau_{F}\sim 1/k_{\perp}$. If we consider a Lorentz boost into a reference frame where $k_{\parallel}\neq 0$, the formation time obtains a dilation factor of approximately $\gamma_{\text{eff}}(k_{\parallel},k_{\perp})\approx k_{\parallel}/k_{\perp}$ for $k_{\parallel}\gg k_{\perp}$.

\begin{figure}[htb]
 \begin{center}
  \includegraphics[height=5.0cm]{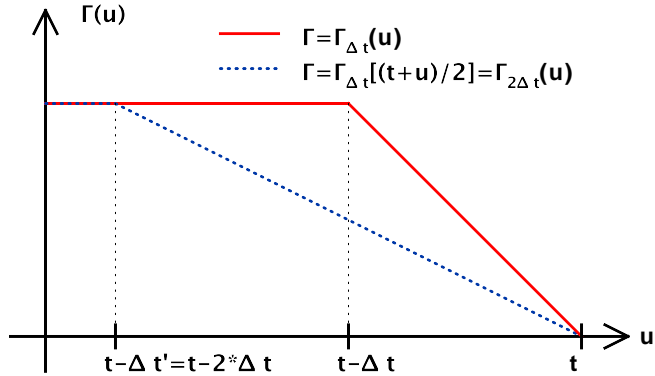}
  \caption{At the time $t$ when the full change is reached, $\Gamma=\Gamma\left(\frac{t+u}{2}\right)$ for
           $\Delta t$ corresponds to $\Gamma=\Gamma(u)$ for $2\Delta t$.}
  \label{fig:compare}
 \end{center}
\end{figure}

We shall briefly explain why the memory times are only half as big for $\Gamma=\Gamma\left(\frac{t+u}{2}\right)$ as they are for $\Gamma=\Gamma(u)$. At the time $t$, when the full change is reached, $\Gamma=\Gamma\left(\frac{t+u}{2}\right)$ is equal to $\Gamma=\Gamma(u)$ for twice the switching time interval $\Delta t$ which can also be inferred from Fig. \ref{fig:compare}. So for $\Gamma=\Gamma\left(\frac{t+u}{2}\right)$, the photon rate after $\Delta t$ has already developed as far as for $\Gamma=\Gamma(u)$ and $\Delta t'=2\Delta t$. So memory effects are only half as big.

\begin{figure}[htb]
 \begin{center}
  \includegraphics[height=4.0cm]{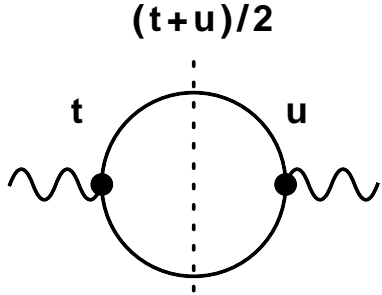}
  \caption{The time dependence of $\Gamma$ enters $i\Pi^{<}_{\mu\nu}(\vec{k},t,u)$ via both time arguments as this expression is the 
           current-current-correlator for the points of time $t$ and $u$.}
  \label{fig:one_loop_cut}
 \end{center}
\end{figure}

Even though both implementations render a causal time evolution, $\Gamma=\Gamma\left(\frac{t+u}{2}\right)$ can nevertheless be considered as the more physical one. As the photon self energy $i\Pi^{<}_{\mu\nu}(\vec{k},t,u)$ is the current-current-correlator for the different points of time $t$ and $u$, one intuitively expects that the change of $\Gamma$ enters $i\Pi^{<}_{\mu\nu}(\vec{k},t,u)$ via both time arguments and not only via one of them (see Fig. \ref{fig:one_loop_cut}). Since the memory times are clearly below 1 fm/c for $\Gamma=\Gamma\left(\frac{t+u}{2}\right)$, our results already indicate that the adjustment process of the radiative behavior of the fermion jet to changes in the medium is almost instantaneous.

Now we turn to the total photon rate and the total radiation power, which are obtained by integrating (\ref{eq:rate_final}) over all photon modes. Since $\Gamma$ does not depend on the photon momentum $\vec{k}$, the total photon rate and the total radiation power are logarithmically and linearly divergent, respectively. This unphysical behavior is cured by introducing a cutoff at $k=p$, preventing the fermion jet from emitting photons with higher energy than its own:
\begin{subequations}
 \label{eq:yield}
 \begin{eqnarray}
  \frac{dn_{\gamma}(t)}{dt} & = & \int_{0}^{p} k^{2}dk\int d\Omega_{k} \ \frac{d^{4}n_{\gamma}(t)}{dt d^{3}k} 
                                  \label{eq:yield_photons} \\
  \frac{dE(t)}{dt}          & = & \int_{0}^{p} k^{2}dk\int d\Omega_{k} \ k\frac{d^{4}n_{\gamma}(t)}{dt d^{3}k}
                                  \label{eq:yield_energy}
 \end{eqnarray}
\end{subequations}

\begin{figure}[htb]
  \hfill
  \begin{minipage}[t]{.45\textwidth}
      \begin{center}
        \includegraphics[height=5.0cm]{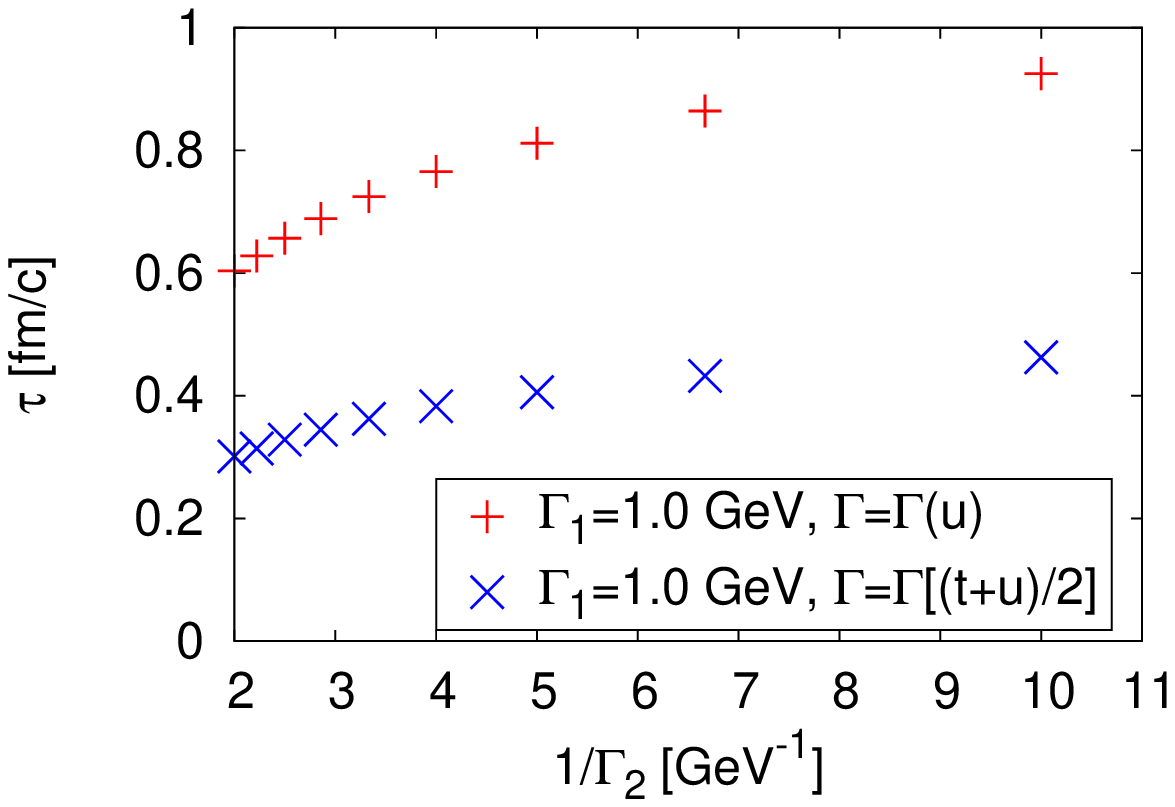}
        \caption{Memory times for the total photon rate}
        \label{fig:mtpn}
      \end{center}
  \end{minipage}
  \hfill
  \begin{minipage}[t]{.45\textwidth}
      \begin{center}
        \includegraphics[height=5.0cm]{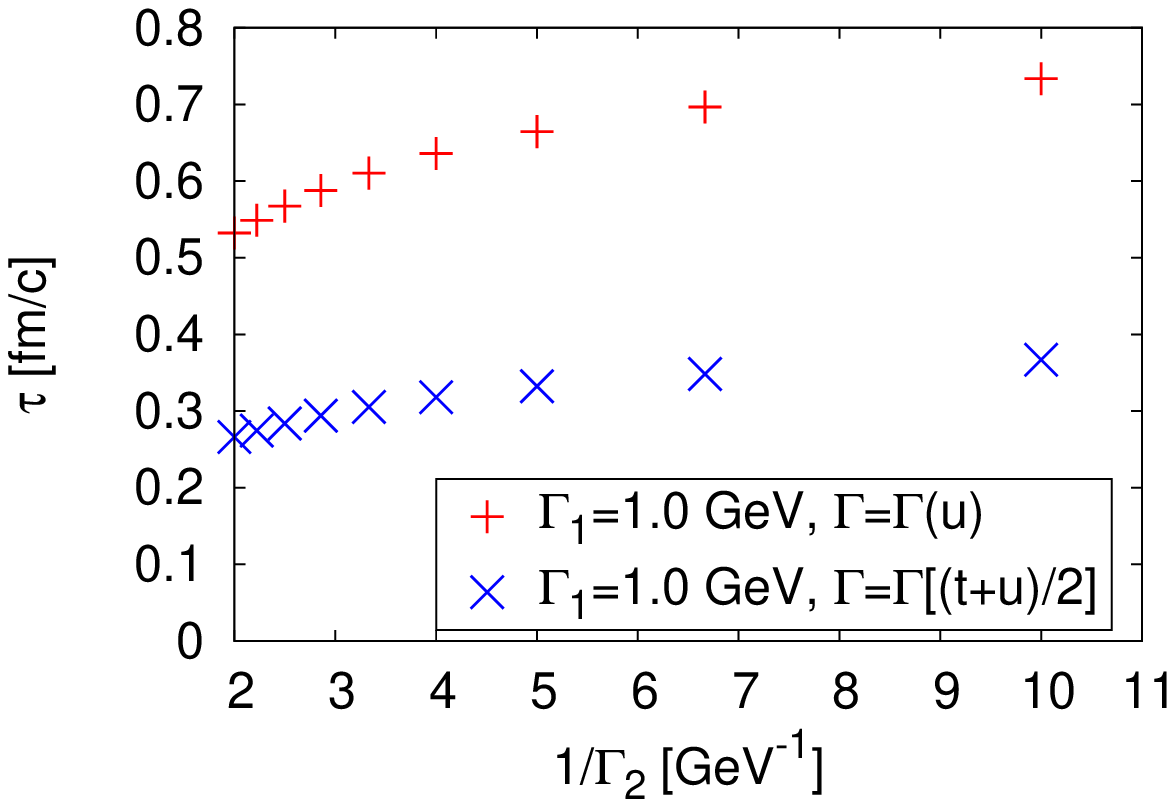}
        \caption{Memory times for the total radiation power}
        \label{fig:mtrp}
      \end{center}
  \end{minipage}
  \hfill
\end{figure}

The memory times for these two quantities are calculated in the same way as for the individual photon modes and shown in Figs. \ref{fig:mtpn} and \ref{fig:mtrp}. Again the memory effects are only half as big for $\Gamma=\Gamma\left(\frac{t+u}{2}\right)$ as they are for $\Gamma=\Gamma(u)$. The overall behavior corresponds to that of the photon modes in the LPM-region as these modes have significantly larger memory times than those in the BH-region. The memory times of the total radiation power are smaller than those of the total photon rate, because soft photons with larger memory times are suppressed by a factor of $k$. Again the adaption of both quantities to the scattering rate $\Gamma$ is almost Markovian ($\tau<1$ fm/c).

\begin{figure}[htb]
  \hfill
  \begin{minipage}[t]{.45\textwidth}
      \begin{center}
        \includegraphics[height=5cm]{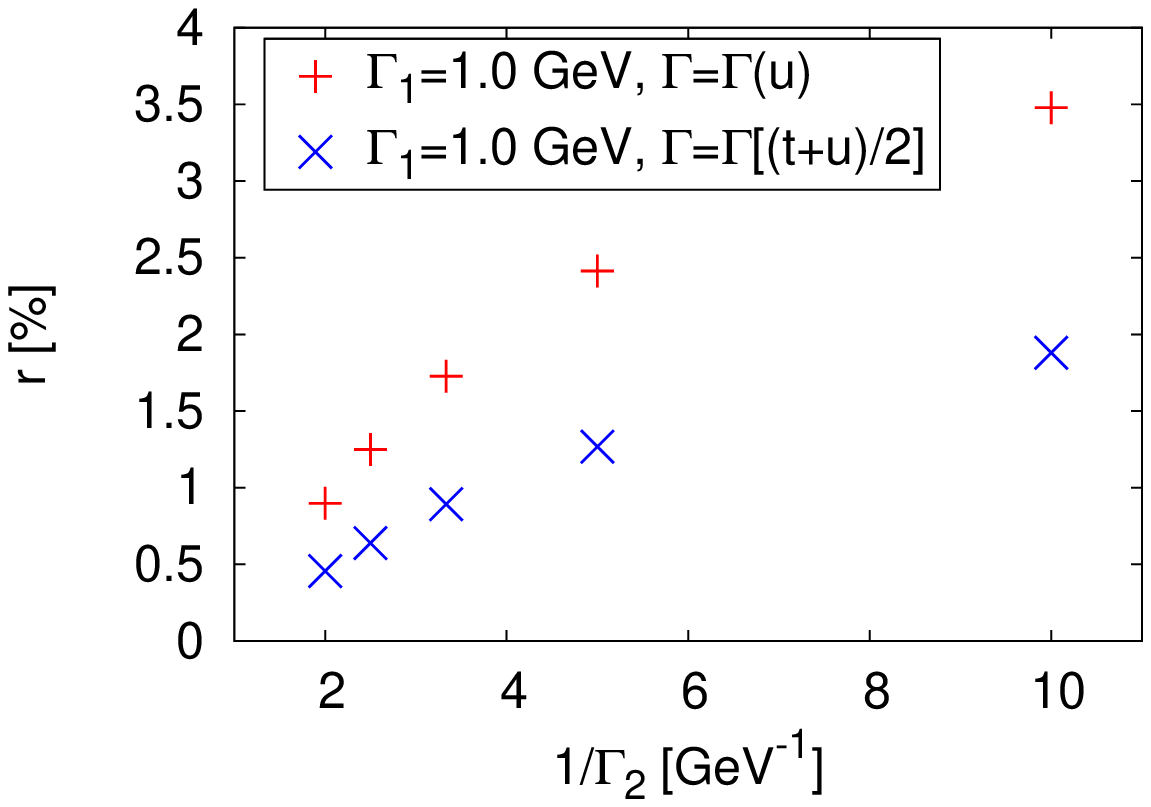}
        \caption{Relative difference for $\Delta t=4$ fm/c}
        \label{fig:rd_4}
      \end{center}
  \end{minipage}
  \hfill
  \begin{minipage}[t]{.45\textwidth}
      \begin{center}
        \includegraphics[height=5cm]{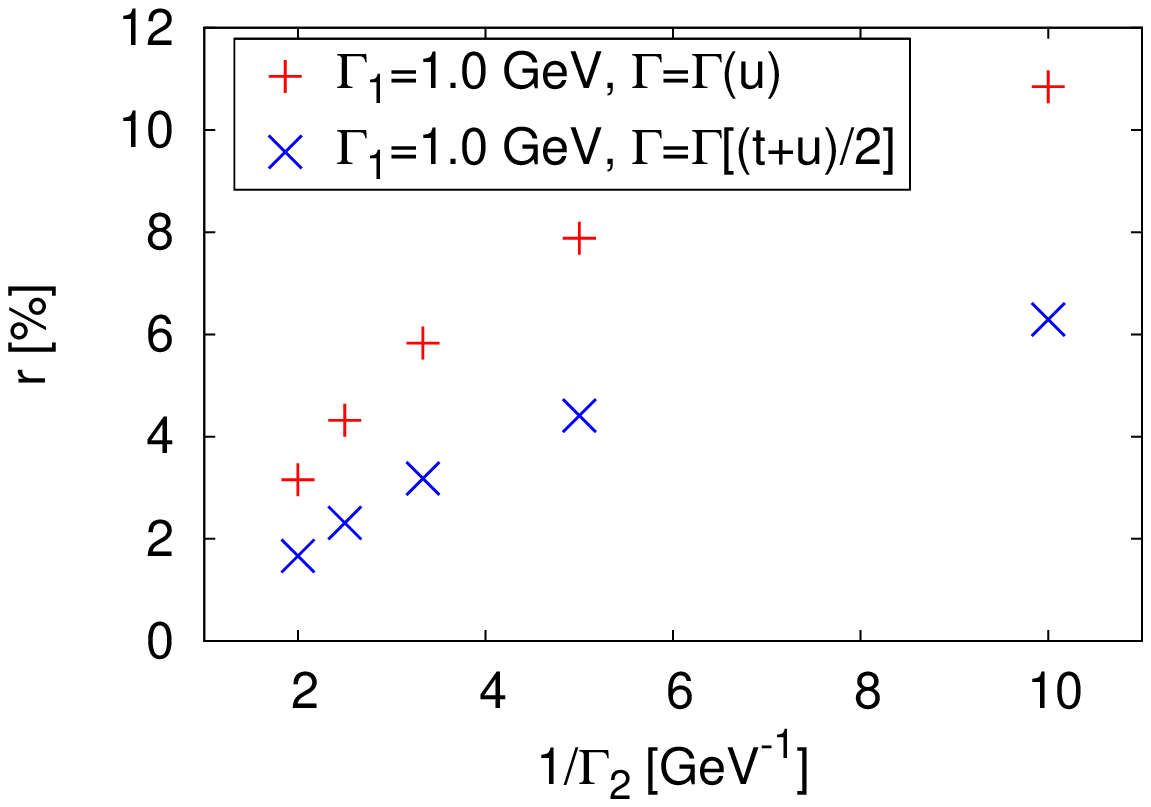}
        \caption{Relative difference for $\Delta t=1$ fm/c}
        \label{fig:rd_1}
      \end{center}
  \end{minipage}
  \hfill
\end{figure}

Finally, we calculate the energy loss of the fermion jet during the time $\Delta t$. A comparison of our result to a quasistatic calculation, as shown in Figs. \ref{fig:rd_4} and \ref{fig:rd_1}, reveals a difference of less than 5 percent for the characteristic expansion time ($\Delta t=4$ fm/c) and scattering rates ($\Gamma_{1}=1.0$ GeV, $\Gamma_{2}=0.3$ GeV) of a quark-gluon plasma. A significant aberration ($\gtrsim 10$ percent) can only be observed for $\Delta\tau=1$ fm/c. This shows again that the adjustment process is close to Markovian.

In the case of $\Delta t=4$ fm/c, we get an energy loss of $-\Delta E=6.32$ GeV which amounts to about one third of the fermion jet's onshell energy. Such a high energy loss is clearly unrealistic for photon emission even though the scattering rate is dominated by strong interaction processes. If we considered gluon emission instead, the electromagnetic coupling constant of $\alpha_{e}\approx1/147$ in (\ref{eq:rate_final_re}) would be replaced by the strong coupling constant of $\alpha_{s}\approx 0.3$ such that the energy loss would increase by a factor of roughly 40 and the fermion jet would be damped off immediately.
\begin{figure}[htb]
  \hfill
  \begin{minipage}[t]{.45\textwidth}
      \begin{center}
        \includegraphics[height=5.0cm]{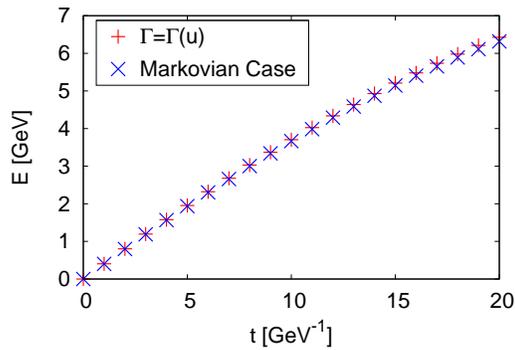}
      \end{center}
  \end{minipage}
  \hfill
  \begin{minipage}[t]{.45\textwidth}
      \begin{center}
        \includegraphics[height=5.0cm]{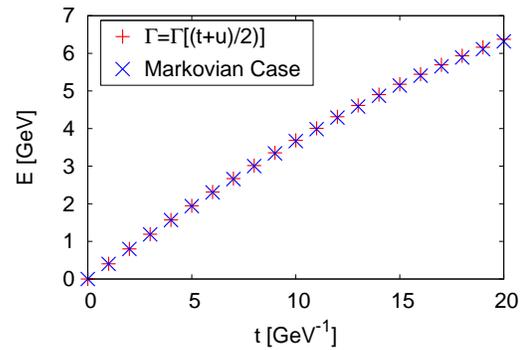}
      \end{center}
  \end{minipage}
  \caption{Energy Loss for $\Delta t=4$ fm/c}
  \label{fig:el_4}
  \hfill
\end{figure}

The main reason for such a high energy loss is the slow convergence of (\ref{eq:rate_final}) for large $k$. Without the cutoff in equations (\ref{eq:yield_photons}) and (\ref{eq:yield_energy}) the total radiation power and thus the energy loss would even be infinite. This behavior results from our ansatz for $S^{<}$ and $S^{>}$ which allows for transitions between arbitrarily offshell energy states which is not physical. In order to see this, we consider a modified ansatz where the fermion can only undergo transitions between energy states which do not differ from the onshell energy by more than $n\Gamma$:
\begin{subequations}
 \label{eq:ansatz_mod}
 \begin{eqnarray}
   S_{F}^{<}(\vec{p}+\vec{k},E) & = & 2\pi i\,\frac{(2\pi)^{3}}{V}\delta^{(3)}(\vec{p}+\vec{k}-\vec{p}')
                                      C(n)\,\Theta(n\Gamma-|E-\varepsilon_{\vec{p}+\vec{k}}|)\,A_{F}(\vec{p}+\vec{k},E) \\
   S_{F}^{>}(\vec{p},E)         & = & -2\pi\,i\Theta(n\Gamma-|E-\varepsilon_{\vec{p}}|)\,A_{F}(\vec{p}+\vec{k})
 \end{eqnarray}
\end{subequations}
Here $n$ can be any positive number and must not necessarily be an integer. The expression $C(n)$ in $S_{F}^{<}(\vec{p}+\vec{k},E+k)$ 
is a normalization factor which is given by $C(n)=\pi/2\arctan(n)$. It is necessary to keep the normalization condition for a single spin-1/2 fermion
\begin{equation}
 -iV\int\frac{d^{3}pdE}{(2\pi)^{4}}Tr\left\lbrace\gamma_{0}S^{<}_{F}(\vec{p},E)\right\rbrace=2 \ .
\end{equation}
The propagator $S^{<}_{F}(\vec{p},E)$ is normalized to $2$ because the spectral function only includes the particle and not the antiparticle component. One easily verifies that with the above prescription one has no more photon emission for $\omega_{k}>2n\Gamma$. Therefore the total radiation power and the total energy loss is tremendously reduced for small $n$ which is shown in Table \ref{tab:res} for the parameters $\Gamma_{1}=1.0$ GeV, $\Gamma_{2}=0.3$ GeV and $\Delta t=4$ fm/c. The case $n=\infty$ corresponds to the original result.
\begin{table}[H]
 \begin{center}
  \begin{tabular}{c|c|c|c|}
    $n$ & $dE/dt$ [GeV$^{2}$] for $\Gamma_{1}=1.0$ GeV & $dE/dt$ [GeV$^{2}$] for $\Gamma_{2}=0.3$ & $-\Delta E$ [GeV] \\
    \hline
     0.5       &  0.017				       & 0.005					  & 0.22              \\
     1         &  0.052				       & 0.016			                  & 0.70              \\
     2         &  0.120                                & 0.038                                    & 1.60              \\
     5         &  0.239                                & 0.080                                    & 3.25              \\
    10         &  0.327                                & 0.114                                    & 4.53              \\
    $\infty$   &  0.407                                & 0.199                                    & 6.32              \\
  \end{tabular}
  \caption{Radiation power and energy loss for different values of $n$}
  \label{tab:res}
 \end{center}
\end{table}

If we compare the energy loss $-\Delta E$ for different values of $n$ and take into account that its value is scaled up by factor of $\alpha_{s}/\alpha_{e}\approx 40$ for the case of gluon emission, we see that our results become more realistic if $n$ is chosen sufficiently small. For the case of $n=0.5$, the energy loss increases to $-\Delta E=8.8$ GeV for gluon emission. This is almost 50 percent of the onshell energy of the fermion jet which from the phenomenological point of view seems to be quite reasonable.

The energy loss in Table \ref{tab:res} has been calculated for the case of a Markovian time evolution. Within the original ansatz, the strong contribution of hard photon modes with small memory times is also the reason for the small overall memory effects. As the emission of hard photons is suppressed by restricting the number of available energy states, we investigate if this procedure also results in larger memory effects. For this purpose, we dynamically calculate the energy loss for $n=1$ and $n=0.5$ in Eqs. (\ref{eq:ansatz_mod}) and again compare our results to a quasi static calculation.
\begin{figure}[htb]
  \hfill
  \begin{minipage}[t]{.45\textwidth}
      \begin{center}
        \includegraphics[height=5cm]{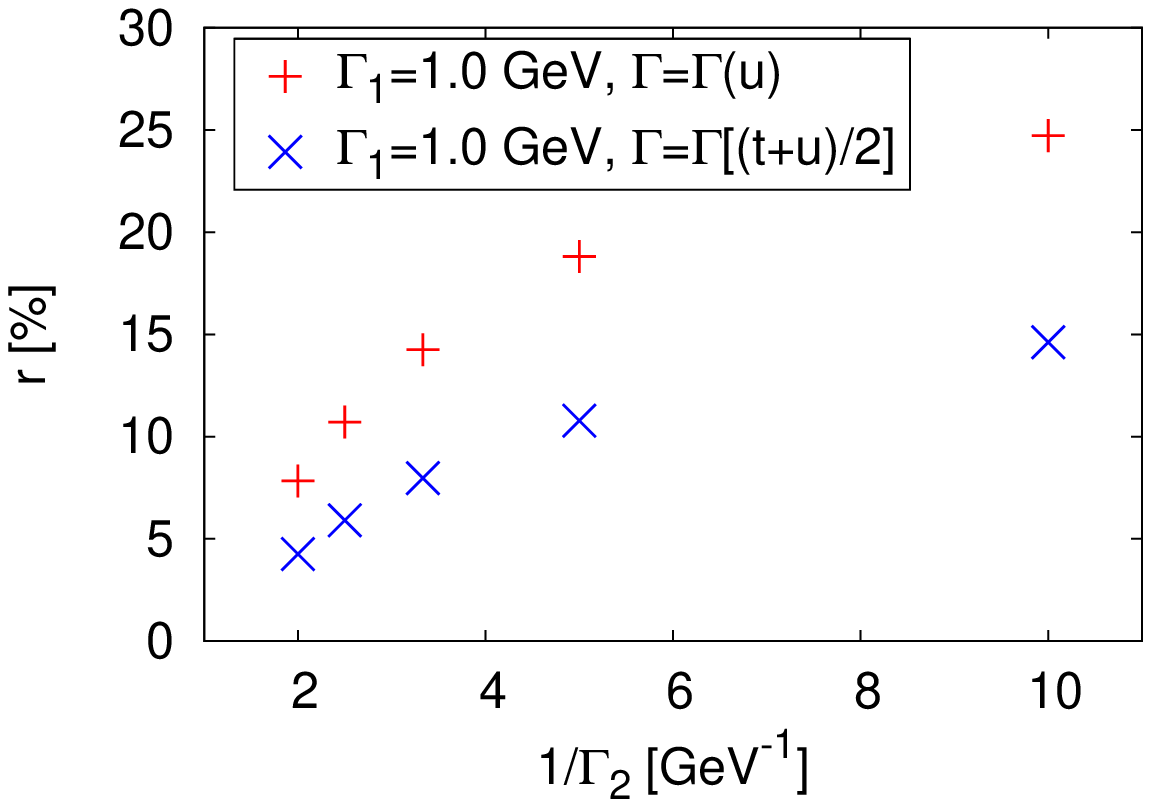}
        \caption{Relative difference for $\Delta t=4$ fm/c and $n=0.5$}
        \label{fig:rd_mod_1}
      \end{center}
  \end{minipage}
  \hfill
  \begin{minipage}[t]{.45\textwidth}
      \begin{center}
        \includegraphics[height=5cm]{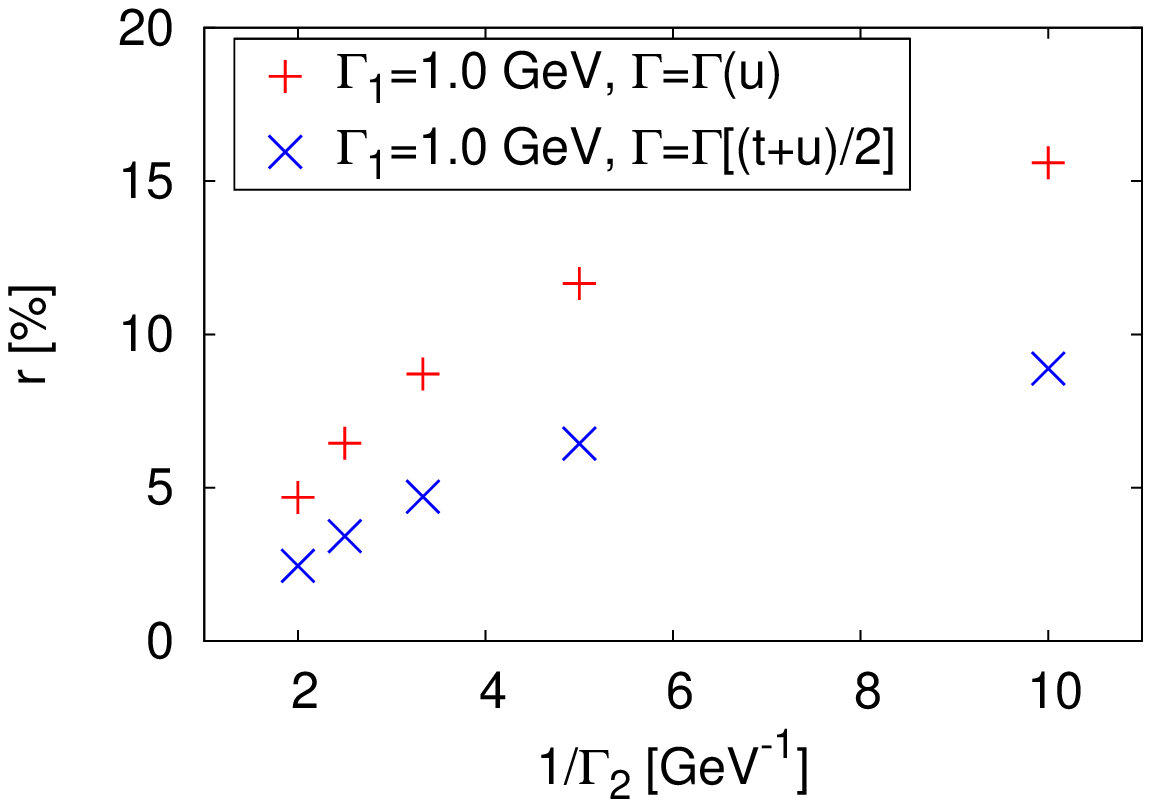}
        \caption{Relative difference for $\Delta t=4$ fm/c and $n=1.0$}
        \label{fig:rd_mod_2}
      \end{center}
  \end{minipage}
  \hfill
\end{figure}

We can infer from Figs. \ref{fig:rd_mod_1} and \ref{fig:rd_mod_2} that the memory effects are clearly larger than those for the original ansatz. We can now observe significant aberrations between the quasi static and the dynamic result for the characteristic expansion time of a quark-gluon plasma ($\Delta t=4$ fm/c). They reach up to 15 percent for n=1.0 and increase even further (up to 25 percent) for n=0.5. So restricting the number of the available initial and final energy states does not only result in a tremendous reduction of the energy loss $-\Delta E$ but memory effects also play a sizable role as the contribution of hard photon modes with small memory time is suppressed.
\begin{figure}[htb]
  \hfill
  \begin{minipage}[t]{.45\textwidth}
      \begin{center}
        \includegraphics[height=5cm]{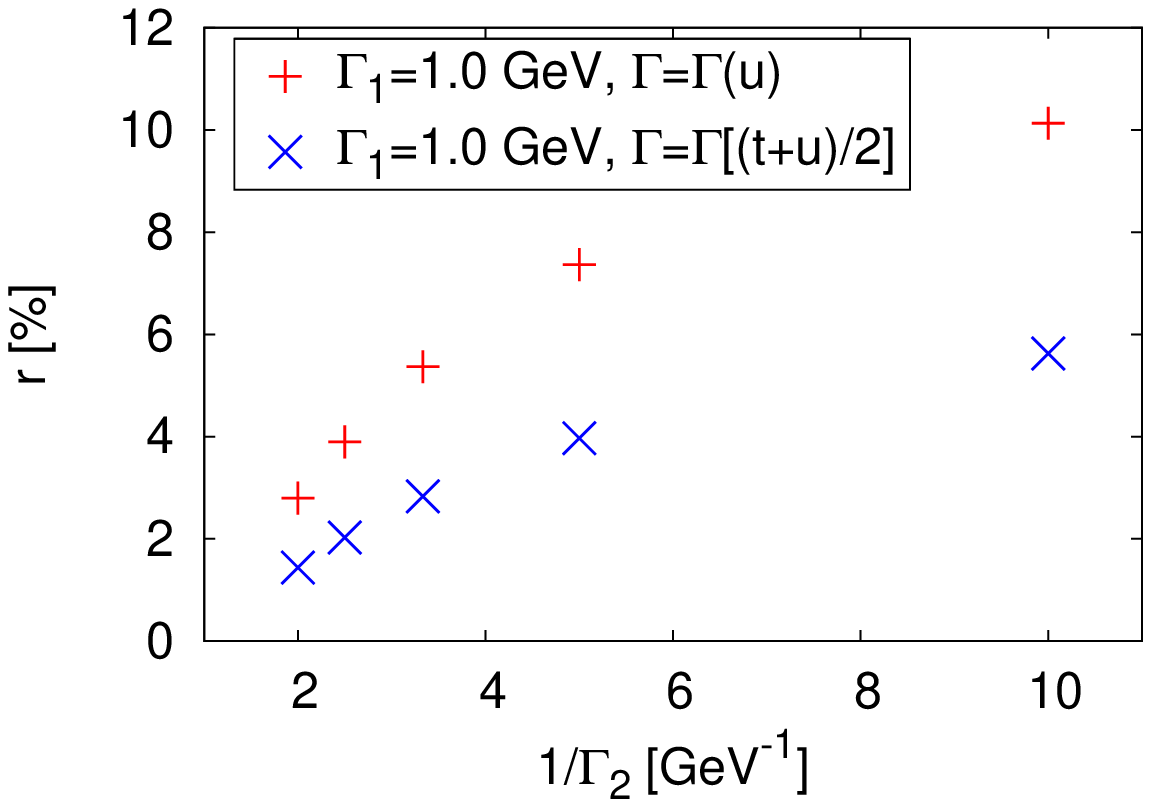}
        \caption{Relative difference for $\Delta t=4$ fm/c and $n=2.0$}
        \label{fig:rd_mod_3}
      \end{center}
  \end{minipage}
  \hfill
  \begin{minipage}[t]{.45\textwidth}
      \begin{center}
        \includegraphics[height=5cm]{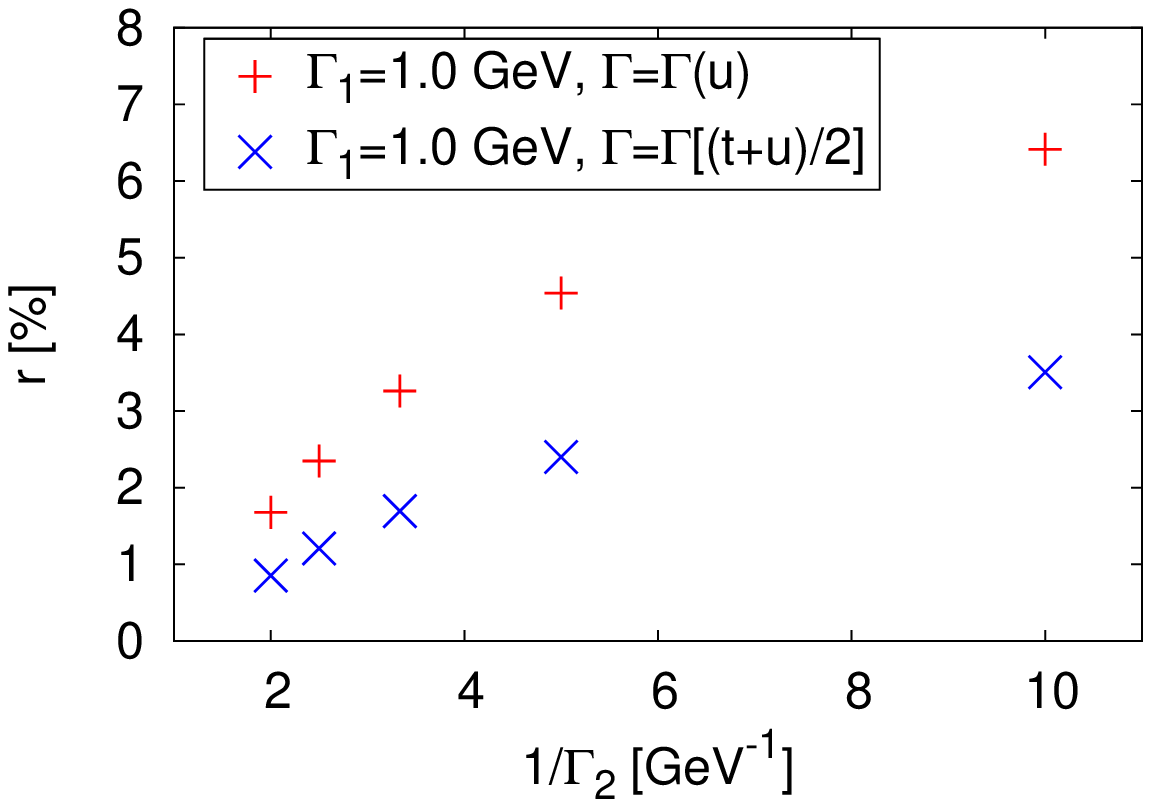}
        \caption{Relative difference for $\Delta t=4$ fm/c and $n=5.0$}
        \label{fig:rd_mod_4}
      \end{center}
  \end{minipage}
  \hfill
\end{figure}

\begin{figure}[htb]
 \begin{center}
  \includegraphics[height=5.0cm]{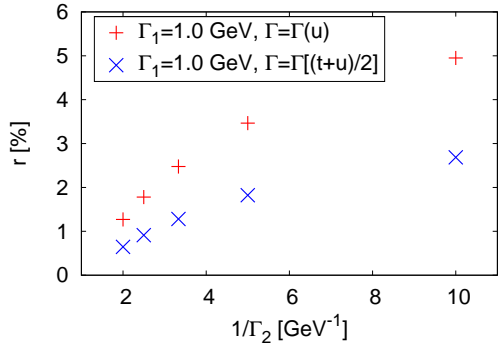}
  \caption{Relative difference for $\Delta t=4$ fm/c and $n=10.0$}
  \label{fig:rd_mod_5}
 \end{center}
\end{figure}

But we have to keep in mind that the size of the memory effects naturally depends on $n$ and that we are principally allowed to choose this parameter freely as the energy loss $-\Delta E$ remains finite for any finite value of it. This in turn results in smaller memory effects for larger values of $n$ which is depicted in Figs. \ref{fig:rd_mod_3}-\ref{fig:rd_mod_5}. 
So getting more reliable results about their relevance for radiative jet energy loss probably needs a full microscopic and quantum field theoretical approach to this phenomenon.

\section{Summary and Outlook}
In this work, we have investigated the electromagnetic radiative energy loss of a fermion jet which moves through the evolving quark-gluon plasma created during heavy-ion collisions at high energies. We focused on whether the radiative behavior of the jet adjusts to the evolving medium instantaneously or if existing memory effects are of importance. For this purpose we have applied the real time formalism of finite temperature field theory, which allows for a full quantum mechanical treatment of such a non-equilibrium phenomenon. Using an effective in-medium propagator for the fermion-jet, we determined the time scales in which the radiative behavior of the jet adjusts to changes in the medium, and also calculated the energy loss of a fermion jet over the time in which the plasma expands and cools down. As a first step, we have restricted ourselves to photon emission.

For the characteristic parameters of a quark-gluon plasma (effective initial and final scattering rates of the jet quark: $\Gamma_{1}=1.0$ GeV, and $\Gamma_{2}=0.3$ GeV and a life time of approximately $\Delta t=4$ fm/c), the memory times for the evolving total photon rate and total radiation power have been found to be in physical terms clearly below 1 fm/c. Furthermore, a comparison of our result for the total energy loss with a quasistatic calculation has revealed a difference of less than 5 percent. This supports the validity of quasi-equilibrium calculations when calculating photon production from a QGP and indicates that a full dynamic treatment is not crucial. The main reason for that are the large scattering rates in a QGP leading to a fast decorrelation of the source particle defining the major time scale for memory effects. If one considers dilepton production from the subsequent hadronic phase where the scattering rates are smaller \cite{SG:2005,Schenke:2006uh}, memory effects play a more important role.

But for completeness, it is important to point that the small memory effects also result from our specific ansatz for the fermion propagators which allows for transitions between arbitrarily offshell energy states. This results in a strong contribution of hard photon modes with small memory times leading to the small overall memory effects mentioned above. As our ansatz requires the introduction of a cutoff in order to keep the total photon rate, the total radiation power and the total energy loss finite, it is strictly speaking not physical. For that reason, we have also considered a modified ansatz where the number of available energy states was restricted to a finite interval around the onshell energy which reduces the emission of hard photons. Besides a strong reduction of the totally emitted energy $-\Delta E$, this also leads to significantly larger memory effects. The magnitude of these effects does, of course, depend on the chosen width of the energy interval. Thus, even though we were able to determine the role of memory effects for all photon modes separately, finding a final answer to the question whether they are relevant for the total radiative jet energy loss might indeed require a full microscopic and quantum mechanical treatment.

In this work, we have restricted our considerations to QED processes. The extension of our investigations to the non-Abelian QCD will bring  complications. Gluons can interact with each other and additional self energy diagrams become relevant. Another important problem in the QCD framework will be that the formation time of the emitted gluon depends on the momentum transferred to the gluon during the rescattering in the medium. This phenomenon has already been investigated in \cite{Baier:1994bd}.

\section{Acknowledgements}
F. M. gratefully acknowledges financial support by the Helmholtz Research School for Quark Matter Studies (H-QM) and from the Helmholtz Graduate School for Hadron and Ion Research (HGS-HIRe for FAIR). B. S. gratefully acknowledges a Richard H. Tomlinson grant by McGill University and support by the Natural Sciences and Engineering Research Council of Canada. This work was (financially) supported by the Helmholtz International
Center for FAIR within the framework of the LOEWE program (Landesoffensive zur Entwicklung Wissenschaftlich-Ökonomischer Exzellenz) launched by the State of Hesse.

\bibliography{photons}
\end{document}